# A REVIEW OF INTERFERENCE REDUCTION IN WIRELESS NETWORKS USING GRAPH COLORING METHODS

Maaly A. Hassan[1] and Andrew Chickadel[2]

[1]Department of Computer Engineering, Islamic University of Gaza, Gaza, Palestine
maaly_awad@hotmail.com
[2]Department of Computer Science, Villanova University, Villanova, USA
andrew.chickadel@villanova.edu

## ABSTRACT

*The interference imposes a significant negative impact on the performance of wireless networks. With the continuous deployment of larger and more sophisticated wireless networks, reducing interference in such networks is quickly being focused upon as a problem in today's world. In this paper we analyze the interference reduction problem from a graph theoretical viewpoint. A graph coloring methods are exploited to model the interference reduction problem. However, additional constraints to graph coloring scenarios that account for various networking conditions result in additional complexity to standard graph coloring. This paper reviews a variety of algorithmic solutions for specific network topologies.*

## 1. INTRODUCTION

One of the main challenges of wireless communication is interference. Unfortunately, research in this area is so young which leads researchers to have different ideas regarding the identification of a universal measure of network interference. According to the Glossary of Telecommunication Terms - Federal Standard 1037C, interference is defined as:

***Interference:*** *A coherent emission having a relatively narrow spectral content, e.g., a radio emission from another transmitter at approximately the same frequency, or having a harmonic frequency approximately the same as another emission of interest to a given recipient, and which impedes reception of the desired signal by the intended recipient.*

Informally speaking, a node *u* may interfere with another node *v* if *u*'s interference range unintentionally covers *v*. Consequently, the amount of interference experienced by a node *v* corresponds to the amount of interference produced by nodes whose transmission range covers *v*. In frequency division multiplexing cellular networks, reducing the amount of interference results in fewer channels which, in turn, can be exploited to increase the bandwidth per frequency channel. In systems using code division multiplexing, small interference helps in coding overhead. In the context of ad hoc and sensor networks, there is an additional motivation for keeping interference low. In these networks consisting of battery driven devices, energy is typically scarce and the frugal usage of it is critical in order to prolong system operability and network lifetime. In addition to enhancing throughput, minimizing interference may help in lowering node energy dissipation by reducing the number of collisions (or the amount of energy spent in an effort of avoiding them) and consequently retransmissions on the media access layer.

                                                                  58

Interference can be reduced by having nodes send with less transmission power. The area covered by the smaller transmission range will contain fewer nodes, yielding less interference. On the other hand, reducing the transmission range has the consequence of communication links being dropped. However, there is surely a limit to how much the transmission power can be decreased. In ad hoc networks, if the node's transmission ranges become too small and too many links are abandoned, the network may become disconnected. Hence, transmission ranges must be assigned to nodes in such a way that the desired global network properties are maintained.

Transmitting nodes influence the ability of other nodes to receive data. A node is not able to receive data from its neighbor if it was interfered by receiving a transmission not intended for it. This mutual disturbance of communication is called interference. Reducing interference in the network leads to fewer collisions and packet retransmissions, which indirectly reduces the power consumption and extends the lifetime of the network. Therefore, reducing the interference is an important goal for wireless networks. The interference imposes a potential negative impact on the performance of a wireless network. In MANETs, each device can selectively decide which device to communicate with either by adjusting its transmission power or its antenna direction. Obviously, keeping relatively limited direct neighbours is helpful to speed up the routing protocols in addition to possibly alleviating the interference among simultaneous transmissions, and also possibly save the energy consumption.

If a network incurs a large interference, either many communication signals sent by nodes will collide, or the network may experience a serious delay at delivering the data for some nodes, and even consume more energy. So, we reach to the conclusion that the interference is a major drawback of wireless networks. The aim of reducing interference is to prevent adjacent or connected nodes, which are linked by radio signals, from receiving and transmitting signals which conflict or blend together. Thus, interference occurs when conflicting transmissions over one radio frequency are received by one or more nodes in a wireless network. This inhibits the ability of the receiver to decipher incoming signals. This concept is illustrated in Figure 1a, which shows a typical situation in which the broadcast areas of nodes A and C overlap in the vicinity of node B, causing B to receive a garbled signal composed of the signals from A and C. In such situations, it is difficult for B not only to decipher simultaneous signals, but also to reliably determine the source of the signal. The problem of reducing interference in arbitrary networks turns out to be very difficult, and for this reason, simpler network layouts have been investigated such as, multi-hop wireless mesh network layouts [1], triangular lattice topologies [9], unit disk graphs [3], hexagonal topologies [8], and other more general topologies [4]. Other key facets of the interference problem in wireless networks specify whether a proposed solution is contrived in a distributed or centralized setting, whether nodes in a given solution are self-aware of their location or whether this assumption is not necessary, and whether or not minimum separation distance between nodes needs to be factored into algorithmic solutions.

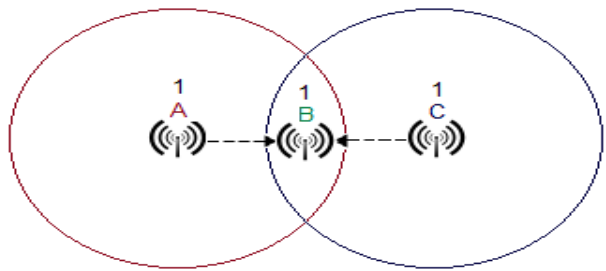

Interference occurs at B

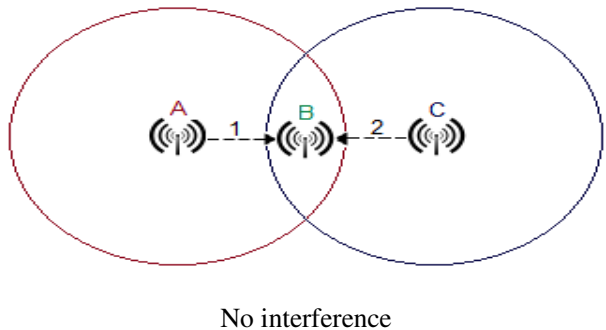

No interference

Figure 1. Interference resolved through channel assignment

Numerous methods for reducing interference exist, such as topology control [10], power control [7], and channel assignment [1, 2, 3, 5, 6, 8, 9]. This paper will focus exclusively on the latter method, which seeks to assign channels of different frequencies to interfering nodes or edges.

In Figure 1a, for instance, simultaneous transmissions of A and C result in interference at B. This problem is resolved in Figure 1b by having nodes A and C transmit over frequencies 1 and 2 respectively, equipping node B with two radios that can transmit and receive over frequencies 1 and 2. Signals from A and C can be demultiplexed, (that is, components of different frequencies can be extracted from one signal) at node B because of differing frequencies, and node B can clearly determine if node A transmits across frequency 1 or if node C transmits across frequency 2. Therefore the intersection node A and node C's broadcast areas no longer results in interference. Through careful assignment of communication channels to nodes in a network, interference could be greatly reduced. It is important to note, however, that the number of radio frequencies is finite, and therefore, the problem of minimizing the number of channels allocated to a specific network is worthy of thorough investigation as well. In some instances, channel overlap is necessary if the number of assigned channels for a network is inadequate to connect all nodes [2].

The rest of the paper is organized as follows. In Section 2, we define the interference problem as a graph coloring problem and discuss two coloring approaches, vertex and edge coloring. Finally a summary of existing results and conclusions are presented in Section 3.

## 2. RELATED WORK

Arunesh Mishra et. al. in [2] propose techniques to improve the usage of wireless spectrum in the context of wireless local area networks (WLANs) using new channel assignment methods among interfering Access Points (APs). They identify new ways of channel re-use that are based on realistic interference scenarios in WLAN environments. In this paper they formulated channel assignment in WLANs as a weighted vertex coloring problem that takes into account realistic channel interference observed in wireless environments, as well as the impact of such interference on wireless users. They proposed two efficient, scalable and fault tolerant distributed algorithms that achieve significantly better performance than the state-of-the-art Least Congested Channel Search (LCCS). Through simulations, they showed that the two techniques achieve up to 45.5% and 56% reduction in interference for sparse and dense topologies respectively with 3 non-overlapping channels. They also show that the techniques effectively use partially overlapping channels to achieve an additional 42% reduction on average for moderately sized networks. They validated these results using experiments on a fully operational in-building wireless testbed network comprising of 20 APs and achieved a 40% reduction using partially overlapping channels. A straightforward extension to this work is to handle co-existing 802.11b/g APs in the same area of coverage. The overlap graph in such scenarios becomes directed in nature as the interference effects become asymmetric (802.11g APs would be more affected than 802.11b). The weights on the edges would reflect a measure of the asymmetric effect of the interference caused by one AP's BSS to another. We leave such extensions as future work. Finally they prove that the weighted graph coloring problem is NP-hard and propose scalable distributed algorithms that achieve significantly better performance than existing techniques for channel assignment.

Mathieu Couture et. al. in [3] present the first location oblivious distributed unit disk graph coloring algorithm having a provable performance ratio of three (i.e. the number of colors used by the algorithm is at most three times the chromatic number of the graph). This is an improvement over the standard sequential coloring algorithm since they present a new lower bound of 10/3 for the worst-case performance ratio of the sequential coloring algorithm. The previous greatest lower bound on the performance ratio of the sequential coloring algorithm was 5/2. However, simulation results showed that this algorithm does not provide a significant improvement over the algorithm which sequentially colors the nodes in an arbitrary order. Simulation results also showed that, in the average case, largest-first (which is also distributed and location oblivious) performs better than the algorithm they proposed. It also performs better than lexicographic coloring, which also has a worst-case performance ratio of at most three. However, no one has shown whether largest-first has a better worst-case performance ratio than five. In fact, it is also an open question whether coloring the nodes of a unit disk graph in an arbitrary order can, on the worst case, use less than five or more than 10/3 times the minimum number of colors that are necessary.

## 3. VERTEX VS. EDGE COLORING

When used without any qualification, a coloring of a graph is almost always a proper vertex coloring, namely a labeling of the graph's vertices with colors such that no two vertices sharing the same edge have the same color. Since a vertex with a loop could never be properly colored, it is understood that graphs in this context are loopless. The terminology of using colors for vertex labels goes back to map coloring. Labels like red and blue are only used when the number of colors is small, and normally it is understood that the labels are drawn from the integers {1,2,...}. A coloring using at most *k* colors is called a (proper) *k*-coloring. The smallest number of colors needed to color a graph G is called its chromatic number, $\chi$ (G). A graph that can be assigned a (proper) *k*-coloring is *k*-colorable, and it is *k*-chromatic if its chromatic number is exactly *k*. A subset of vertices assigned to the same color is called a color class; every such class forms an independent set. Thus, a *k*-coloring is the same as a partition of the vertex set into *k* independent sets, and the terms *k*-partite and *k*-colorable have the same meaning.

An edge coloring of a graph is a proper coloring of the edges, meaning an assignment of colors to edges so that no vertex is incident to two edges of the same color. An edge coloring with *k* colors is called a *k*-edge-coloring and is equivalent to the problem of partitioning the edge set into *k* matchings. The smallest number of colors needed for an edge coloring of a graph G is the chromatic index, or edge chromatic number, $\chi'$ (G). A Tait coloring is a 3-edge coloring of a cubic graph. The four color theorem is equivalent to the assertion that every planar cubic bridgeless graph admits a Tait coloring.

Interference in wireless networks is commonly represented as an interference graph in which edges represent potential interference between the endpoint nodes.  The problem of reducing collisions and signal interference is modeled as a coloring problem on the interference graph. Nodes of different colors in the graph will be assigned separate channel of radio frequency. Efficient coloring algorithms will lead to an effective channel selection method that lowers the wireless interference. The graph coloring problem's application is relevant because preventing vertices from connecting via radio frequency with other (conflicting) vertices is the quintessential task in reducing interference.  Graph coloring algorithms are also reliable in that they are mathematically provable. Channel assignment, however, adds algorithmic complexity to standalone graph coloring problems, according to Khanna and Kumaran [8].   Channel assignment is indeed what ties the graph coloring problem together with the problem of reducing interference in wireless networks. Figure 2 illustrates a proper vertex coloring of the Petersen graph with 3 colors, the minimum number possible. Figure 4 also shows 3-colored graph which can be colored in 12 different ways

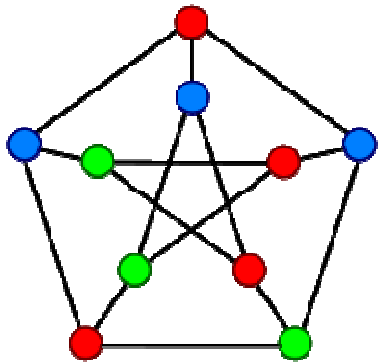

Figure 2. A proper vertex coloring of the Petersen graph with 3 colors, the minimum number possible

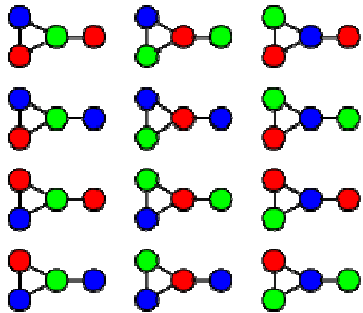

Figure 3. This graph can be 3-colored in 12 different ways

There is ongoing research explicitly investigating vertex-coloring and edge-coloring graph methods that address the channel assignment problem for wireless networks [1, 2, 3, 5, 6, 8, 9]. The order in which colors or channels are assigned and to what nodes differs greatly from one solution to another.  The creation of such an order is also quite varied among current proposed solutions.  Weighted coloring is one method that deals exclusively with assigning channels based on need to alleviate interference within the network.

### 3.1. Weighted Coloring

One variation of the graph coloring problem involves assigning weights to the interference graph. A weighted coloring implementation addresses interference chiefly in areas of greatest need. Once these problem areas are discovered, channel re-assignment can alleviate signal collision. McDiarmid and Reed [9] assign weights based on bandwidth demands, whereas Arbaugh et al. [2] assign weights that indicate the degree of channel interference between two nodes. McDiarmid and Reed [9] reduce interference by assigning to each node a number of colors equal to its weighted bandwidth demand value. Arbaugh et al. [2] devise algorithms, which they call Hminmax and Hsum, which greedily choose the frequency at each node which will locally at that node result in the greatest reduction in interference. McDiarmid and Reed [9] state that there is future work in determining an improved ratio for large demands. Arbaugh et al. [2] state, however, that for their algorithms Hminmax and Hsum, wireless-b and wireless-g environments were not studied in a mixed setting, and this remains an open issue. Figure 4 depicts a network where there are only three non-overlapping channels for four wireless access points labeled AP4, AP5, AP6, and AP7. Each access point can broadcast far enough to reach each wireless device attached to it. In this instance, two access points will need to share a communication channel [2].

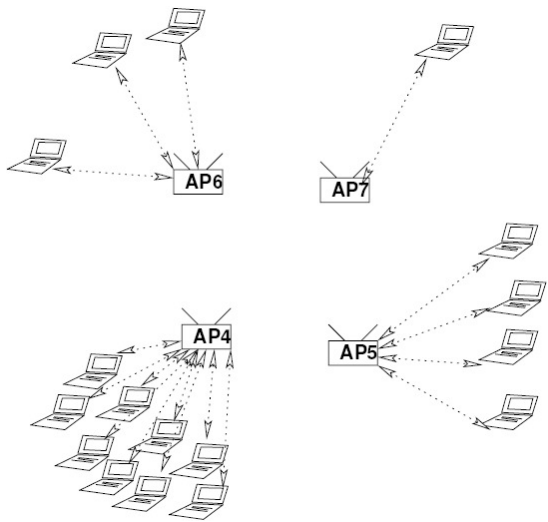

Figure 4. All Access Points (AP) are in interference range of each other [2]

A greedy weighted algorithm would assign AP4 and AP5 their own channel to broadcast over, because they have more devices connected than AP6 and AP7 do. AP6 and AP7 therefore will have to share one channel and endure the possibility of interference to spare the majority from the threat of interference. Provided that AP6 and AP7 are sufficiently far from each other, sharing one communication channel may result in no interference. This idea of channel reuse has been explored by several researchers aiming to reduce not only interference, but the number of communication channels as well.

### 3.2. Channel Re-use

Channel re-use involves imposing a minimum distance between two nodes operating at the same radio frequency. Re-using channels not only curtails interference but also improves overall efficiency in terms of channel use. Bertossi et al. [5] define the channel assignment problem with separation (CAPS) in order to incorporate such a re-use distance in channel assignment, resulting in a more efficient use of the given set of radio frequencies. An algorithm is presented by which a node, knowing its relative position in the network, can compute its channel assignment, with a specified re-use distance, in constant time for all network graph types except for binary trees, which require logarithmic time [5]. Bertossi et al. [5] however, identify an open issue exists in determining an optimal solution for general graphs and arbitrary channel re-use distance. An alternative for computing minimum distance would be to determine the minimum and maximum number of channels to use in a network.

### 3.3. Radio Frequency (RF) Spectrum

Investigating strict upper and lower bounds for the number of channels used in a given wireless network is of importance to the interference reduction problem. Khanna and Kumaran [8] define what they call the wireless spectrum estimation problem, where a node is assigned the smallest number of frequencies over which to broadcast and receive, such that interfering nodes do not share the same frequency. Bertossi et al. [5] later explore channel re-use governed by a minimum re-use distance rather than static assignment as delineated by upper and lower bounds (see Section 3.2). Additionally, the ability to compute minimum distance based on relative position is not always a necessity.

### 3.4. Location-Oblivious

Location-oblivious networks do not rely on each node knowing its relative geometric position. According to Barbeau et al. [3], location-aware networks, on the other hand, might require a Global Positioning System (GPS) in order to calculate nodes' relative positions. Therefore, location-oblivious networks are preferable. Barbeau et al. [3] introduce a distributed, location-oblivious unit disk graph coloring algorithm unlike Bertossi et al. [5] whose algorithms require nodes to be self-aware of their location. However, the algorithm presented by Barbeau et al. did not out-perform an algorithm that executes arbitrary coloring of a network graph [3]. Similarly to location-oblivious algorithms, dynamic channel assignment [2] does not require a node to know its relative position.

### 3.5. Dynamic Channel Assignment

Dynamic channel assignment can reduce interference in networks where the topology changes often and dramatically. Dynamic channel assignment involves continuously monitoring interference and re-assigning channels appropriately. Arbaugh et al. [2] employ a form of dynamic assignment in their multi-step, greedy approach algorithms Hminmax and Hsum that re-assign channels only to nodes experiencing the greatest interference (mentioned in Section 3.1). Almeroth et al. [1] introduce dynamic assignment in their algorithm, Breadth First Search - Channel Assignment (BFS-CA). BFS-CA utilizes breadth-first search to determine nodes of greatest connectivity which typically are subject to the greatest interference. Channels are re-assigned on-the-fly, and connectivity is maintained through a secure "default channel" [1]. Figure 4[1] below illustrates why a default channel is vital to the structure of the network.

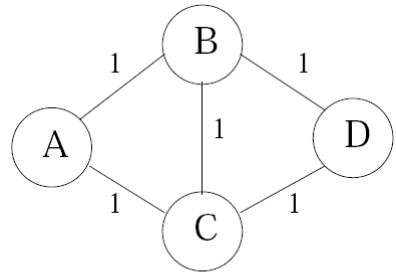

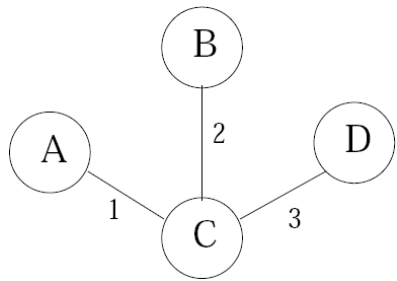

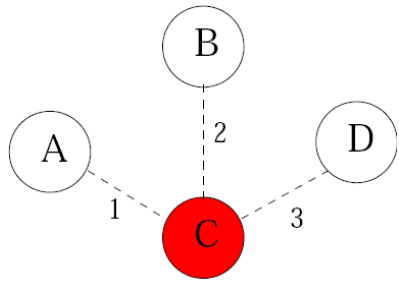

Figure 5. Network topology with varying channel assignments [1]

Figure 5(a) depicts a sample network topology where all nodes broadcast over one frequency. Suppose node C is selected for channel re-assignment via the addition of two new radios that broadcast and receive over channels 2 and 3. Figure 5(b) shows the result of this re-assignment. Through this process of eliminating the threat of interference, nodes A, B, and D lose direct connection to each other and instead must communicate using one-hop over node C. However, further complications arise with this topology change. Figure 5(c) shows the event of node C failing. The dotted lines represent broken connections. The consequence of this is that nodes A, B and D will not be able to communicate, even though they are within communication range of one another. For all changes in topology, node failure is a real threat in breaking communication. With a default channel in place, however, a severed connection due to node failure can be overcome with minimal downtime [1]. In addition to how BFS-CA [1] will perform in very dense topologies, addressing just how long the downtime due to node failure will last is an open issue, and, in fact, graph coloring algorithms in general are works-in-progress because of their NP-hard and NP-complete algorithmic complexities.

## 4. CONCLUSIONS

The dynamic algorithm, BFS-CA [1], is one of the best algorithms for use in today's ever-changing wireless network topologies. It is the most implementation-ready compared to other graph coloring algorithms. BFS-CA was also shown to have a significant improvement over static assignment of channels [1]. The weighted Hminmax and Hsum [2] algorithms, despite resorting to greedy implementations, have achieved over a 40% average reduction in interference over one "state-of-the-art" method [2]. The McDiarmid and Reed bandwidth-based weighted algorithms [9] bring together several novel ideas, however they seem difficult to eventually implement.

## REFERENCES

[1] K.C. Almeroth, E. M. Belding, M. M. Buddhikot, K. N. Ramachandran, (2006) "Interference-Aware Channel Assignment in Multi-Radio Wireless Mesh Networks,"

[2] W. Arbaugh, S. Banerjee, A. Mishra, (2005) "Weighted Coloring Based Channel Assignment for WLANs," *ACM SIGMOBILE Mobile Computing and Communications Review*, vol. 9, no. 3, pp. 19-31.

[3] M. Barbeau, P. Bose, P. Carmi, M. Couture. E. Kranakis, (2006) "Location Oblivious Distributed Unit Disk Graph Coloring" *School of Computer Science Carleton University* pp. 1-20.

[4] R. Battiti, A. A. Bertossi, M. A. Bonuccelli, (1999) "Assigning Codes in Wireless Networks: Bounds and Scaling Properties," *Wireless Networks*, pp. 195-209.

[5] A. A. Bertossi, C. M. Pinotti, R. B. Tan, (2000) "Efficient use of radio spectrum in wireless networks with channel separation between close stations," *Workshop on Discrete Algorithms and Methods for MOBILE Computing and Communications and Proceedings of the 4th International Workshop on Discrete Algorithms and Methods for Mobile Computing and Communications*, pp. 18-27.

[6] T. Chiueh, K. Gopalan, A. Raniwala, (2004) "Centralized Channel Assignment and Routing Algorithms for Multi-Channel Wireless Mesh Networks," *ACM SIGMOBILE Mobile Computing and Communications Review*, vol. 8 no. 2, pp. 50-65.

[7] D. J. Goodman, N. B. Mandayam, C. U. Saraydar, (2002) "Efficient Power Control Via Pricing in Wireless Data Networks," *IEEE Transactions on Communications*, vol. 50, issue 2, pp. 291-303.

[8] S. Khanna, K. Kumaran, (1998) "On Wireless Spectrum Estimation and Generalized Graph Coloring," *INFOCOM '98. Seventeenth Annual Joint Conference of the IEEE Computer and Communications Societies. Proceedings. IEEE*, vol. 3, pp. 1273-1283.

[9] C. McDiarmid, B. Reed, (2000) "Channel Assignment and Weighted Coloring," *Networks*, vol. 36, no. 2, pp. 114-117.

[10] R. Ramanathan, R. Rosales-Hain, (2000) "Topology Control of Multihop Wireless Networks Using Transmit Power Adjustment," *INFOCOM 2000. Nineteenth Annual Joint Conference of the IEEE Computer and Communications Societies. Proceedings. IEEE*, vol. 2, pp. 404-413.

[11] Scalable Network Technologies, Inc. QualNet Network Simulator. 2007. http://www.scalable-networks.com.

[12]     Andrew Muir, J.J. Garcia-Luna-Aceves, (1998) "A channel access protocol for multihop wireless networks with multiple channels,"

[13]     P. Karn. Maca (1990) "a new channel access method for packet radio," *In Proceedings of ARRL/CRRL Amateur Radio Computer Networking Conference*.

**Authors**

Maaly Awad Hassan obtained her Master and Bachelor degrees in Computer Engineering from Islamic University of Gaza IUG. Her area of interests includes Interference Reduction in Mobile Ad Hoc and Sensor Networks, Internet Technology, Network Security, Image Processing, Distributed Computing, Interference Reduction, Topology Control, MANETS, and Wireless Sensor Networks.

Andrew Chickadel obtained his Bachelor degree in Computer Science (Minor: Cognitive Science) from Villanova University. He has professional experience in the following areas within JPMorgan Chase Card Services Technology: Application development, Mainframe architecture, Quality assurance, Global markets, Process management, Project management, Project lifecycle, and Corporate project delivery process.